%% file: main.tex
\documentclass[journal,twoside,web]{ieeecolor}

\usepackage{generic}
\usepackage{graphicx}
\usepackage{amsmath,amssymb,amsfonts}
\usepackage{makecell,multirow}
\usepackage{booktabs}
\usepackage{cite}
\usepackage{hyperref}
\usepackage{xcolor}

\def\FIGUREPATH{figure/}
\def\BibTeX{{\rm B\kern-.05em{\sc i\kern-.025em b}\kern-.08em
    T\kern-.1667em\lower.7ex\hbox{E}\kern-.125emX}}
\title{Toward Real-World Voice Disorder Classification}
\author{Heng-Cheng Kuo, Yu-Peng Hsieh, Huan-Hsin Tseng, Chi-Te Wang, \\Shih-Hau Fang, and Yu Tsao, \IEEEmembership{Senior Member, IEEE}
\thanks{Manuscript received xxxx; revised xxxx, 2022; accepted xxxxx. This work was supported by Far Eatsern Memorial Hospital under Grant FEMH2021-C-029. The review of this paper was arranged by Editor Paolo
xxxxx. (Corresponding author: Yu Tsao.)}
\thanks{Heng-Cheng Kuo, Yu-Peng Hsieh, Huan-Hsin Tseng, and Yu Tsao are with the Research Center for Information Technology Innovation, Academia Sinica, Taipei 11529, Taiwan (e-mail:{hckuo145, htseng, yu.tsao}@citi.sinica.edu.tw).}
\thanks{Y. Tsao is also a jointly appointed professor of the Department of Electrical Engineering, Chung Yuan Christian University, Taoyuan 32023,
Taiwan}
\thanks{Chi-Te Wang is with the Department of Electrical Engineering, Yuan Ze University, Taoyuan 320, Taiwan, and also with the Department of
Otolaryngology Head and Neck Surgery, Far Eastern Memorial Hospital, New Taipei 220, Taiwan (e-mail: drwangct@gmail.com).}
\thanks{Shih-Hau Fang is with the
Department of Electrical Engineering, Yuan Ze University, Taoyuan 320,
Taiwan (e-mail: shfang@saturn.yzu.edu.tw).}
\thanks{Copyright (c) 2021 IEEE. Personal use of this material is permitted. However, permission to use this material for any other purposes must be obtained from the IEEE by sending an email to pubs-permissions@ieee.org.}}

\begin{document}

\maketitle
\input{0-abstract}
\input{1-introduction}
\input{2-related}
\input{3-methodology}
\input{4-experiment}
\input{5-future}
\input{6-conclusion}
\input{appendix}

\bibliographystyle{IEEEtran}
\bibliography{reference}

\end{document}

%% file: 0-abstract.tex
\begin{abstract}
\textbf{Objective:}
Voice disorders significantly compromise individuals' ability to speak in their daily lives.
Without early diagnosis and treatment, these disorders may deteriorate drastically.
Thus, automatic classification systems at home are desirable for people who are inaccessible to clinical disease assessments.
However, the performance of such systems may be weakened due to the constrained resources and domain mismatch between the clinical data and noisy real-world data.
\textbf{Methods:}
This study develops a compact and domain-robust voice disorder classification system to identify the utterances of health, neoplasm, and benign structural diseases.
Our proposed system utilizes a feature extractor model composed of factorized convolutional neural networks and subsequently deploys domain adversarial training to reconcile the domain mismatch by extracting domain-invariant features.
\textbf{Results:}
The results show that the unweighted average recall in the noisy real-world domain improved by 13\% and remained at 80\% in the clinic domain with only slight degradation.
The domain mismatch was effectively eliminated.
Moreover, the proposed system reduced the usage of both memory and computation by over 73.9\%.
\textbf{Conclusion:}
By deploying factorized convolutional neural networks and domain adversarial training, domain-invariant features can be derived for voice disorder classification with limited resources.
The promising results confirm that the proposed system can significantly reduce resource consumption and improve classification accuracy by considering the domain mismatch.
\textbf{Significance:}
To the best of our knowledge, this is the first study that jointly considers real-world model compression and noise-robustness issues in voice disorder classification.
The proposed system is intended for application to embedded systems with limited resources.
\end{abstract}


%% file: 1-introduction.tex
\section{Introduction}

Early epidemiological studies reported varying estimates of the prevalence of voice disorders, ranging from 0.65\% to 15\%~\cite{laguaite1972adult, morley1952ten}.
A later report estimated the prevalence among the US to be approximately 3\% to 9\%~\cite{ramig1998treatment}.
More recently, a regional telephone survey of 1326 random subjects revealed a current voice disorders prevalence of 6.6\% and a lifetime prevalence of 29.9\% in adults aged less or equal to 65 years~\cite{roy2005voice}.
Another study based on primary care physicians also demonstrated similar results on the lifetime prevalence (4.3\% to 29.1\%) and current prevalence (7.5\%) of voice disorders~\cite{cohen2010self}.
Two large-scale claims data-based epidemiological studies revealed that the prevalence rate of voice disorders ranges from 0.26\% to 0.98\%~\cite{best2011prevalence, cohen2012prevalence}.
All studies have indicated that the overall prevalence of voice disorders is quite alarming.
For such diseases, accurate diagnosis requires experienced specialists and expensive equipment.
Patients without health insurance or other medical resources would face a few months of waiting times for specialist appointments.
Appropriate and instant disease assessment may be inaccessible to people in need.
Therefore, this study proposes a non-invasive self-screening classification system that allows individuals to diagnose pathological voices (health, neoplasm, and benign structural diseases) at home to help schedule the priority of medical resource allocation.
For example, if a patient is diagnosed with neoplasm using the proposed classification system, his/her appointment can be brought forward to reduce the waiting time.
On the other hand, if the self-screening result shows healthy, the user can avoid the risk of infection while traveling to the hospital and the waste of medical resources, especially during epidemics.

In recent decades, several non-invasive screening methods have been proposed, and their potential to identify samples of pathological voices has been demonstrated~\cite{dibazar2002feature, henriquez2009characterization, vaziri2010pathological, 9783160}.
Furthermore, pathological voices always accompany changes in the voice quality~\cite{time_course, illa2021pathological, halpern2021objective, unger2015noninvasive}.
Thus, in previous research, acoustic features, such as Mel frequency cepstral coefficients (MFCCs)~\cite{fraile2009automatic, costa2008pathological, harar2020towards}, glottal features~\cite{umapathy2005discrimination, putzer2021electroglottographic}, and gammatone spectral latitude (GTSL)~\cite{zhou2022gammatone}, were used as inputs of classic machine learning (ML) classifiers, including Gaussian mixture models (GMM)~\cite{godino2006dimensionality}, support vector machine (SVM)~\cite{arjmandi2012optimum, markaki2011voice, hammami2016pathological, verde2018voice, pishgar2018pathological, arias2011combining, arias2010automatic}, and k-nearest neighbors (KNN)~\cite{dahmani2020recurrence, basalamah2023highly}.
On the other hand, various neural network (NN)-based models also verified the reliability of deep learning (DL)~\cite{wu2018convolutional, gupta2018voice, fang2019detection, hung2022using, lee2021experimental, ariyanti2021ensemble, david2021deep}.
Automatic speech recognition systems were used to assess voice disorders as well~\cite{lee2016automatic, liu2019acoustical}.
Besides, some studies have added auxiliary inputs, for example, medical records~\cite{fang2019combining} and the GRBAS scale~\cite{kojima2021objective}, to help classify pathological voices.
Based on the promising performance under ideal conditions, Hsu \textit{et al.}~\cite{hsu2018robustness} further addressed the channel effect due to hardware variation; Fan \textit{et al.}~\cite{fan2021class} and Jinyang \textit{et al.}~\cite{jinyang2021pathological} investigated the sample imbalance between voice disorders. 
With the development of the Internet of Things (IoT), IoT and cloud technology has also been applied to voice pathology monitoring~\cite{muhammad2017smart, muhammad2018edge, hossain2019smart, amin2019cognitive}.
In addition to the above research, the FEMH Challenge~\cite{ramalingam2018ieee}, an international competition held by the IEEE Big Data conference in Seattle in 2018, provided a common dataset and evaluation metrics for the development of voice disorder classification systems~\cite{pham2018diagnosing, grzywalski2018parameterization, bhat2018femh, degila2018ucd, arias2018byovoz, islam2018transfer}. 
The dataset was published by the Far Eastern Memorial Hospital (FEMH), Taiwan.

Nevertheless, current NN-based solutions are not optimal for practical applications due to two main challenges.
First, to achieve state-of-the-art performance, large and deep model structures of neural networks are typically designed.
However, the limited memory and computational resources of embedded systems provide little room for models to increase the number of parameters.
Second, the domain mismatch between the standardized training data and testing data acquired from real-world scenarios substantially degrades the accuracy.

Because a large model requires a huge memory capacity and computational resources, there are typically three common solutions to achieve real-time processing on embedded systems: quantization techniques, knowledge distillation, and factorized convolutional neural networks (CNNs).
Quantization is a technique that replaces the arithmetic of 32-bit floating points with that of integers~\cite{jacob2018quantization} or powers of two~\cite{lin2021seofp}, allowing for a much lower latency on commonly available hardware.
Moreover, owing to the limited number of quantized values, representations with even lower bits can be applied to further reduce the usage of memory.
An alternative solution, knowledge distillation is the process of transferring knowledge from a large network, particularly for an ensemble of models, to a small one~\cite{hinton2015distilling}.
The outputs generated by the cumbersome but well-trained network act as additional labels for the distilled network.
By imitating the behavior of a large network, a distilled network can achieve better performance than using only true labels.
Finally, the standard CNN can be regarded as a combination of spatial convolution in each channel (also called \textit{intra-channel convolution}) and linear projection across channels simultaneously~\cite{wang2017factorized}.
The factorized CNN was developed to rearrange the spatial convolution~\cite{romera2017erfnet} or address these two parts separately~\cite{howard2017mobilenets} to reduce memory and computation.

Another issue in the real-world scenarios is the \textit{domain mismatch}.
Although joining abundant labeled data from different environments is likely to improve the generalizability of models, it is not feasible to prepare rather diverse and out-of-clinic data in the biomedical area due to extreme time consumption.
Additionally, labeling such data requires a strong professional background, which further increases the difficulty.
As a result, data collected in the laboratory or clinics are the few (and sometimes the only) labeled data available.
Our intention is to focus on unsupervised domain adaptation, which requires no labeled data from the real-world domain but labeled data from the clinic domain during the training process.
In general, labeled data defined as the \textit{source domain} have one probability distribution, while unlabeled data, which we intend to adapt to, called the \textit{target domain}, have another.
There are a few methods for generalizing a model to an unseen target domain, including the Generative Adversarial Network (GAN)-based and discrepancy-based methods.
In the GAN-based method, a \textit{generator} generates plausible target domain data with labels from the given source data~\cite{bousmalis2017unsupervised}, where the plausibility is governed by a \textit{discriminator}.
Subsequently, the labeled data from the source domain together with the generated target data are utilized for the main task training.
Thus, information from both the source and the target domains is revealed to the main task model.
Another discrepancy-based method intends to learn extracted features by minimizing the gap between the probability distributions of the source and the target domains so that a well-trained model can be directly applied to the target domain to fit our purpose.
To derive such domain-invariant features, predefined statistics~\cite{tzeng2014deep, lin2021unsupervised, hu2021variational} or domain classifier~\cite{ganin2016domain, pinheiro2018unsupervised} are introduced to assess the discrepancy of probability distributions between the domains.

In this study, we propose a new voice disorder classification system customized for embedded devices, which adapts to daily noisy environments simultaneously.
The proposed model consists of factorized CNNs to obtain compact architecture and is augmented with a domain adversarial training (DAT) module during training to equip it with the ability to operate in noisy environments.
The results showed that the unweighted average recall (UAR) in the noisy real-world domain improved by 13\%, while that in the clinic domain remained at 80\% with only slight degradation.
In addition, the numbers of parameters and Multiply–Accumulate Operations (MACs) were significantly reduced by 73.9\% and 77.0\%, respectively.

The remainder of this paper is organized as follows: In Section~\ref{sec:related}, the related works, including MobileNet and DAT, are reviewed.
Section~\ref{sec:method} introduces the proposed robust voice disorder classification system.
The experimental results and an ablation study are presented in Section~\ref{sec:experiment}.
Section~\ref{sec:future} claims our plans for the future work.
Finally, the conclusions are presented in Section~\ref{sec:conclusion}.

%% file: 2-related.tex
\section{Related Works}\label{sec:related}

\subsection{Model compression}
To achieve a high penetration rate of self-diagnosis at home, classification systems will confront the limits of memory and computational resources on embedded devices.
Thus, in addition to accuracy, the model size and computational cost are also prior considerations. 

In speech signal processing, filter-based conversions, such as Mel-spectrograms and MFCCs, are typically applied to temporal signals.
Therefore, CNN-based models can effectively extract the local pathological characteristics from the (2D image-like) converted inputs.
From this perspective, the factorized CNNs are suitable for reducing the difficulties.

An Efficient Residual Factorized Network (ERFNet)~\cite{romera2017erfnet} rearranges spatial convolutions to achieve lower resource usage.
It factorizes a 3$\times$3 convolution into the union of perpendicular 3$\times$1 and 1$\times$3 convolutions, and residual connections are used to improve the training efficiency while retaining remarkable accuracy under the constrained scenario.

Another factorized method, separable convolution, proposed in MobileNet~\cite{howard2017mobilenets}, is aimed at mobile and embedded applications.
A \textit{separable} convolutional layer factorizes a standard convolutional layer into a depth-wise convolutional layer and a point-wise convolutional layer.
A standard convolution deals with the spatial convolution in each channel and the linear transformation across channels simultaneously, whereas a separable convolution splits this operation into two stages.
Specifically, in the first stage, the depth-wise convolution applies a single filter to each input channel, and in the second stage, the point-wise convolution (simply a 1$\times$1 convolution) then performs a linear projection on the previous depth-wise convolution outputs.
This separate consideration of the relationship in each channel and the relationship between channels drastically reduces the computation and model size.
Two variants, MobileNetV2~\cite{sandler2018mobilenetv2} and MobileNetV3~\cite{howard2019searching}, were proposed to further improve the accuracy and reduce the latency on the successful base of MobileNet.

\subsection{Unsupervised domain adaptation}\label{subsec:dat}
Compared to undisturbed clinics or studios where pathological voices are recorded, background noise is inevitable in daily life where our system is aimed for application.
Moreover, there are very few annotated data in out-of-clinic scenarios available, since it is hard to perform standard and unified experiments for the general public without the assistance of experienced specialists.
Thus, dealing with domain mismatches between the labeled source data and unlabeled target data poses a challenge.
In general, the distributions of these two domains are expected to be similar but not exactly coincident.
In fact, they are required to be "similar" by nature due to the same learning task.
However, slight differences are inevitable between the ages or genders of the subjects, the environments where the data are generated, etc.
The existence of these differences causes performance degradation, especially in data-driven neural networks.
Our purpose is to rectify the data deviation.
Because the GAN-based method~\cite{bousmalis2017unsupervised}, which aims to generate target domain data with labels, requires a large amount of training data, it is not favorable for each situation.
Therefore, the discrepancy-based method is more feasible.

A classification model consists of two parts: a feature extractor and a label predictor.
The feature extractor is designed to extract useful information from the input; subsequently, the label predictor utilizes the extracted features for classification.
Several domain adaptation techniques typically rely on a feature extractor deriving features invariant across domains, \textit{e.g.,} ignoring the background noise or the difference between recording devices \cite{hsu2018robustness}, so that a model can generalize on the target domain while preserving a low risk of misclassification on the source domain~\cite{redko2020survey}.
If the extracted features are perplexing across domains at all times, those features are considered \textit{domain-invariant} in this study.
Based on this idea, statistical techniques or an NN-based domain classifier are introduced to assess the domain invariability of the extracted features.
In the former, Maximum Mean Discrepancy (MMD)~\cite{tzeng2014deep, zhang2022pathological} and Optimal Transport (OT)~\cite{lin2021unsupervised} serve as loss functions to calculate the distance between the probability distributions of the extracted features across domains to be minimized together with classification losses.
In the latter, the extracted features are adversarially trained to perplex the domain classifier and remain high prediction accuracy, where DAT is one of the most popular algorithms augmented with an NN-based domain classifier.

To fool the domain classifier, DAT instructs the feature extractor to update in the opposite direction of minimizing the domain classification loss. For this purpose, the study of DAT introduced a novel gradient reversal layer (GRL) glued by two functions $R$ and $\widetilde{R}$ at different stages, requiring no parameters such that:
\begin{equation}
\begin{aligned}
&R(\mathbf{z}) = \mathbf{z} \quad &\text{(forward propagation)}\\
&\widetilde{R}(\mathbf{z}) = -\mathbf{z} \,\, \Leftrightarrow \,\, \nabla_{\mathbf{z}} \widetilde{R} = - \mathbf{I} \quad &\text{(backward propagation)}
\end{aligned}
\end{equation}
where $\mathbf{z}$ and $\mathbf{I}$ denote the input and identity matrix, respectively.
It should be noted that in a general layer with forward function $\mathbf{z} \mapsto f(\mathbf{z})$, the backward propagation naturally has the derivative $\mathbf{z} \mapsto \nabla_{\mathbf{z}} f$ from the same $f$.
The GRL deliberately splits the forward and backward function into two to achieve the designated purpose.
As such, the GRL acts as an identity function during the forward propagation, but multiplies the gradient by $-1$ during back propagation.
Owing to the GRL, the DAT algorithm can be implemented on any existing ML package with little effort.

%% file: 3-methodology.tex
\section{Methodology}\label{sec:method}

\subsection{Proposed method}
We propose a system for voice disorder classification consisting of separable convolutional layers equipped with a DAT architecture.
Our backbone model replaces the standard deep CNN-based convolutional layers with separable convolutional layers.
The standard CNN-based model referenced ensures the performance and the efficiency with reduced computation.
The experiments in Section~\ref{sec:experiment} verify that the performance of the proposed method is comparable to that of the referenced standard CNN-based method.

We note that the associated training process of DAT is given by the  min-max algorithm:
\begin{align}
\arg\max_{\theta_{d}}\min_{\theta_{f}, \theta{y}} \,\, &\mathbb{E}_{x \sim D_{s}}[L_{y}] + \mathbb{E}_{x \sim D_{s} \times D_{t}}[-\lambda L_{d}]\label{eq:objective_func}\\
L_{y} &= -\rm{log}P(\mathbf{y}|\mathbf{x}, \theta_{f}, \theta_{y})\label{eq:label_loss}\\
L_{d} &= -\rm{log}P(\mathbf{d}|\mathbf{x}, \theta_{f}, \theta_{d})\label{eq:domain_loss}
\end{align}
where a data point \{\emph{input, label, domain}\} is denoted by $\{\mathbf{x}, \mathbf{y}, \mathbf{d} \}$, and $\theta_{f}$, $\theta_{y}$, $\theta_{d}$ denote the parameters of the feature extractor, label predictor and domain classifier, respectively.
$D_{s}$ and $D_{t}$ denote the source domain and the target domain data distribution; 
$L_{y}$ represents the cross-entropy loss of the labels, and $L_{d}$ is that of the domains, where $\lambda \geq 0$ is the coefficient that regularizes the loss in Eq.~(\ref{eq:objective_func}).
To improve the performance of the main task, the classification of pathological voices in this work, the feature extractor and the label predictor shall jointly minimize $L_{y}$.
Due to the unsupervised domain adaptation scheme, $L_{y}$ can only be computed on the source domain data.
On the other hand, the domain classifier minimizes $L_{d}$ to enhance the ability to discriminate domains, whereas the feature extractor maximizes $L_{d}$ to obtain the opposite gradients.
For simplicity, we substitute $-L_{d}$ for $L_{d}$ in Eq.~(\ref{eq:objective_func}) and invert minimization and maximization operations correspondingly.

\begin{figure}[htb]
    \centering
    \includegraphics[width=\columnwidth]{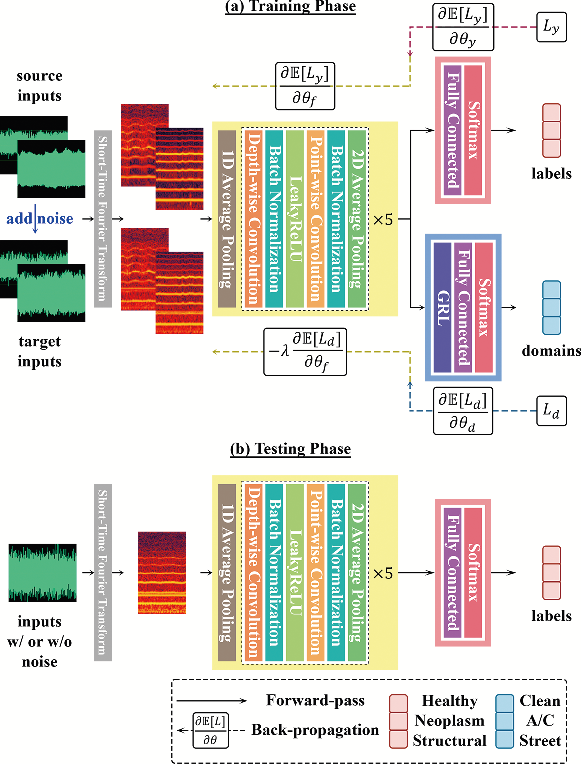}
    \caption{The feature extractor consists of 5 separable convolutional blocks; The disease predictor and domain classifier are simply combinations of a fully connected layer and a softmax layer.}
    \label{fig:flowchart}
\end{figure}

To realize the adversarial min-max Eq.~(\ref{eq:objective_func}), we first formulate the updates of parameters $\theta_{f}$, $\theta_{y}$ and $\theta_{d}$ via the gradient descent \cite{ruder2016overview} as follows:
\begin{align}
\theta_{f} &\longleftarrow \theta_{f} - \alpha \left( \frac{\partial \mathbb{E}[L_{y}]}{\partial \theta_{f}} - \lambda \frac{\partial \mathbb{E}[L_{d}]}{\partial \theta_{f}} \right)  \label{eq:extractor_grad}\\
\theta_{y} &\longleftarrow \theta_{y} - \alpha \,  \frac{\partial \mathbb{E}[L_{y}]}{\partial \theta_{y}} \label{eq:predictor_grad}\\
\theta_{d} &\longleftarrow \theta_{d} - \alpha \, \lambda \, \frac{\partial \mathbb{E}[L_{d}]}{\partial \theta_{d}}\label{eq:classifier_grad}
\end{align}
where $\alpha$ is the learning rate.
The flowchart of the training and testing phases is shown in Fig.~\ref{fig:flowchart}.
Thus, the training phase in Fig.~\ref{fig:flowchart}(a) illustrates the gradient descent process.
The updated formula in Eq.~(\ref{eq:extractor_grad}) and (\ref{eq:classifier_grad}) are observed to have opposite signs with respect to the gradient of $L_{d}$ to conform to the GRL~\cite{ganin2016domain}.

To provide more details, we will elaborate on the flowchart shown in Fig.~\ref{fig:flowchart}.
In this study, the source domain is defined as the clean recording environment like a clinic, whereas the target domain is the environment with noises at home.
During the training phase, the source domain utterances recorded in the clinic are augmented with noises of \textit{air conditions (A/Cs)} and \textit{streets} to synthesize the target domain data and then both are transformed into Log Power Spectrums (LPSs).
The details of the data preprocessing are clarified in the next section.
Next, the lightweight backbone model learns to identify utterances of health, neoplasm, and benign structural diseases from both clean and noisy environments through the DAT technique.
However, real-world utterances with or without noises can be directly used to diagnose diseases through the trained model during the testing phase.
Compared to the high UAR in the source domain, our system only suffers slight degradation of the UAR in the target domain.

\subsection{Model Architecture}\label{subsec:architecture}
The proposed voice disorder classification system is shown in Fig.~\ref{fig:flowchart}.
The feature extractor consists of separable convolutional layers receiving inputs of size (127, 251) with the first and second dimensions denoting the \textit{frame length} and the \textit{frequency basis} of LPSs, respectively.
First, the 1D average-pooling layer (with a kernel size of 2 and stride of 2) reduces the input size along the dimension of frequency to be (127, 126), leaving the frame length unchanged.
Subsequently, the downsampled inputs are extended with a dimension of channels to be (127, 126, 1).
After the first 1D average-pooling layer, five identically separable convolutional blocks follow, each of which comprises a depth-wise convolutional layer (with akernel size of (3, 3, $C_{\text{i}}$) and stride of 1), a point-wise convolutional layer (with a kernel size of (1, 1, $C_{\text{i}}$, 16) and stride of 1), and a 2D average-pooling layer (with a kernel size 2 and stride of 2).
Here, $C_{\text{i}}$ is set to 1 in the first block and $16$ in the rest.
Batch normalization \cite{ioffe2015batch} is applied after all convolutional layers (depth-wise and point-wise), while LeakyReLUs \cite{xu2015empirical} (with a negative slope of 0.2) are inserted after the depth-wise convolutional layers only.
The final output of the feature extractor, viewed as a 1D vector of dimension 256, is regarded as the extracted feature carrying domain-invariant information to pass on to the next disease predictor and the domain classifier.
The disease predictor is a fully connected layer of matrix size (256, $k$) with a softmax layer concatenated right after to predict the probability between the $k$ distinct diseases, in our case, $k=3$ (health, neoplasm, and benign structural diseases).
The domain classifier is similar to the disease predictor, as a three-class classification task, only with the augmentation of a GRL ahead.
It classifies the domains into clean, A/C, and street.
Specifically, \textit{the two types of noises, A/C and street, are annotated separately for the domain classifier}.
The ADAM~\cite{kingma2014adam} optimizer with a learning rate of 0.001 and the regularization coefficient $\lambda = 0.5$ are used throughout the experiments unless otherwise specified.

\subsection{Memory Usage and Computational Cost}
\begin{figure}[htb]
    \centering
    \includegraphics[width=\columnwidth]{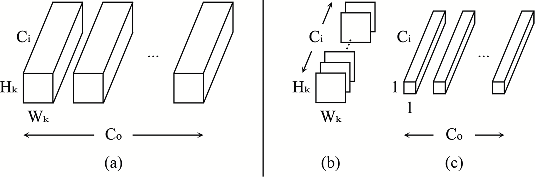}
    \caption{(a) Standard convolutional filters are factorized into (b) depth-wise convolutional filters for intra-channel convolution and (c) point-wise convolutional filters for cross-channel projection.
    }
    \label{fig:sepconv}
\end{figure}

First, we derive the capability of the separable convolutional layer.
Consider a general 3D input $\mathbf{I} \in \mathbb{R}^{W_{i} \times H_{i} \times C_{i}}$, an output  $\mathbf{O} \in \mathbb{R}^{W_{o} \times H_{o} \times C_{o}}$ and a corresponding convolution kernel $\mathbf{K} \in \mathbb{R}^{W_{k} \times H_{k} \times C_{i} \times C_{o}}$ in a standard convolutional layer, where $W_{i}$, $H_{i}$ and $C_{i}$ denote the width, height and number of channels of an input feature respectively;
similarly $W_{o}$, $H_{o}$, $C_{o}$, $W_{k}$, $H_{k}$ denote those of an output feature and a convolution kernel.
In the following, a filter is defined by a group of kernel parameters decomposed along the dimension of output channels.
Therefore, $\mathbf{K}$ is regarded as $C_{o}$ filters of size $W_{k} \times H_{k} \times C_{i}$.

Because a standard convolutional layer is parameterized by its convolution kernel $\mathbf{K}$ (Fig.~\ref{fig:sepconv}(a)), we can directly derive the number of parameters in a single layer:
\begin{equation}\label{eq:n_standard}
W_{k} \cdot H_{k} \cdot C_{i} \cdot C_{o}
\end{equation}
However, due to only intra-channel convolutions, a convolution kernel $\mathbf{\widehat{K}}$ (Fig.~\ref{fig:sepconv}(b)) of the depth-wise convolutional layer comprised of $C_{i}$ filters with a size of $W_{k} \times H_{k}$.
Moreover, the number of channels for the output features remains $C_{i}$.
Therefore, a point-wise convolutional layer combining the information between channels and mapping to the desired shape is essential.
Since we focus on the convolutions across channels, the width and height of a point-wise convolution kernel $\mathbf{\dot{K}}$ (Fig.~\ref{fig:sepconv}(c)) are set to $1$, forming a 1$\times$1 convolutional layer.
In total, the number of parameters of a separable convolutional layer is:
\begin{equation}\label{eq:n_factoried}
W_{k} \cdot H_{k} \cdot C_{i} + 1 \cdot 1 \cdot C_{i} \cdot C_{o}
\end{equation}
By Eq.~(\ref{eq:n_standard}) and (\ref{eq:n_factoried}), the reduction ratio of the model size is:
\begin{equation}\label{eq:size_ratio}
\frac{W_{k} \cdot H_{k} \cdot C_{i} + 1 \cdot 1 \cdot C_{i} \cdot C_{o}}{W_{k} \cdot H_{k} \cdot C_{i} \cdot C_{o}} = \frac{1}{C_{o}} + \frac{1}{W_{k} \cdot H_{k}}
\end{equation}

Next, we discuss the reduction in computation.
The computational cost is dominated by multiplications of floating points, so we analyze the number of multiplications in convolutional layers.
Because each element in the output feature is the dot product of the specific filter and part of the input feature, the number of multiplications is the filter size multiplied by the output size.
The point-wise convolutions merely affect the dimension of channels, and hence the output feature of a depth-wise convolutional layer sizes $W_{o} \times H_{o} \times C_{i}$ after intra-channel convolutions.
The following are computational costs for each type of convolution layer:
\begin{equation}\label{eq:cost_all}
\begin{aligned}
\text{Standard:} \quad &(W_{k} \cdot H_{k} \cdot C_{i}) \cdot (W_{o} \cdot H_{o} \cdot C_{o}) \\
\text{Depth-wise:} \quad &(W_{k} \cdot H_{k}) \cdot (W_{o} \cdot H_{o} \cdot C_{i})\\
\text{Point-wise:} \quad &(1 \cdot 1 \cdot C_{i}) \cdot (W_{o} \cdot H_{o} \cdot C_{o})
\end{aligned}
\end{equation}
The computational reduction ratio is then:
\begin{multline}\label{eq:cost_ratio}
\frac{(W_{k} \cdot H_{k}) \cdot (W_{o} \cdot H_{o} \cdot C_{i}) + (1 \cdot 1 \cdot C_{i}) \cdot (W_{o} \cdot H_{o} \cdot C_{o})}{(W_{k} \cdot H_{k} \cdot C_{i}) \cdot (W_{o} \cdot H_{o} \cdot C_{o})}\\
= \frac{1}{C_{o}} + \frac{1}{W_{k} \cdot H_{k}}
\end{multline}

On the other hand, one important reason to use DAT as the domain adaptation method in our system is that it does not increase any memory load or computational cost in the testing phase.
As shown in the comparison of Fig.~\ref{fig:flowchart}(a) and Fig.~\ref{fig:flowchart}(b).
During the training phase, the feature extractor collaborates with the domain classifier to jointly learn the domain-invariant features.
However, in the testing phase, the disease predictor utilizes the well-trained features from both domains to diagnose diseases without the interference of the domain classifier.
Thus, the number of parameters remains unchanged regardless of whether the DAT technique is used.

Therefore, Eq.~(\ref{eq:size_ratio}) and Eq.~(\ref{eq:cost_ratio}) reveal that the entire memory usage and the computational cost can both be significantly reduced by over $73.9\%$ in our design. 
Additional experimental details are provided in Section~\ref{subsec:exp_all}.

%% file: 4-experiment.tex
\section{Experiments}\label{sec:experiment}

\begin{figure*}[htb]
    \centering
    \includegraphics[width=\linewidth]{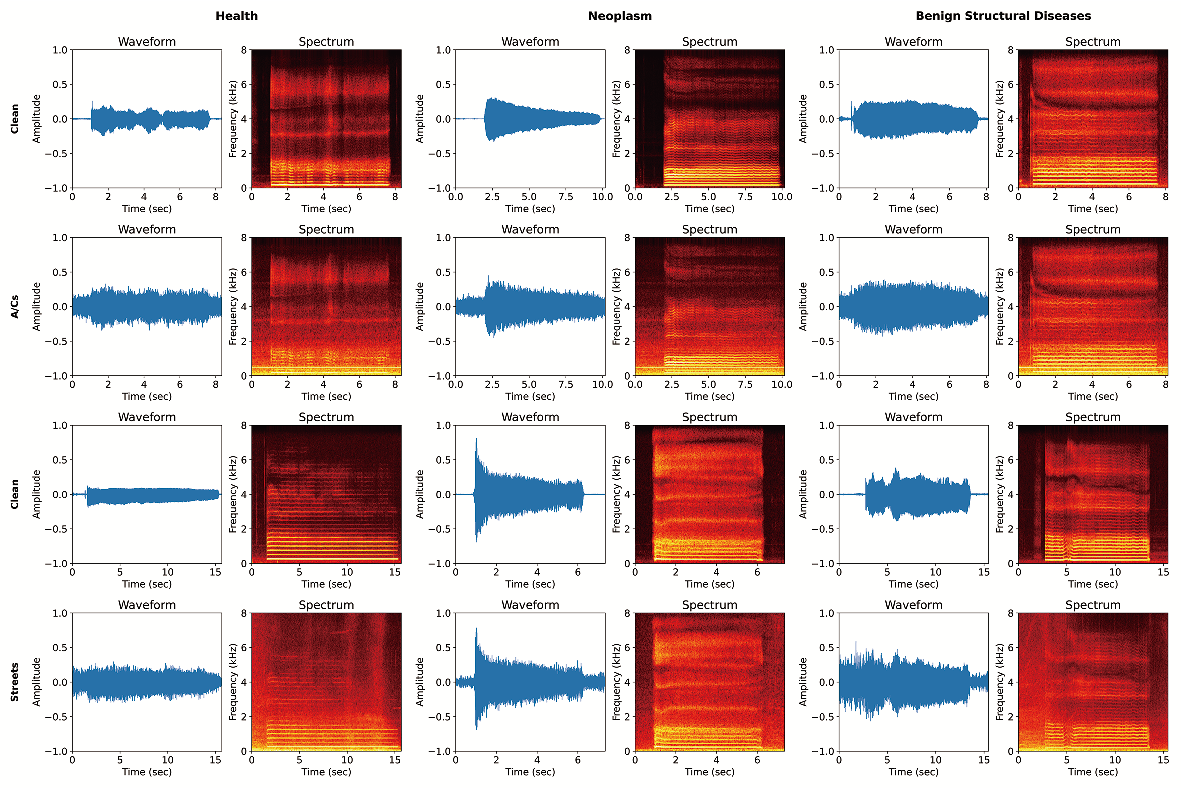}
    \caption{Waveform and spectrogram plots of health, neoplasm, and benign structural diseases speech samples (the vowel /:a/ sound); the first and third rows list the plots of clean utterances, and the second and fourth rows list the plots of noisy utterances corrupted by noises of A/Cs and streets, respectively.}
    \label{fig:input}
\end{figure*}

\subsection{Dataset and Preprocessing}
The voice samples were collected from the Far Eastern Memorial Hospital (FEMH) using a unidirectional microphone and a digital amplifier (CSL model 4150B, Kay Pentax).
All participants were asked to sustain the vowel /:a/ for at least 2 seconds during recordings. 
The sampling rate was 44.1 kHz with a 16-bit resolution and the data were saved in an uncompressed wave format.

A total of 523 voice samples were recorded in a voice clinic as the source domain, containing 108 healthy voices, 112 glottic neoplasms and 303 benign structural diseases (\textit{i.e.,} vocal nodules, polyps and cysts).
Another 30 voice samples, 10 of each category, were synthesized with various noises of A/Cs and streets as the target domain data, where the labels were not provided during the training phase, for unsupervised domain adaptation.
In this study, a 10-times 5-fold cross-validation approach was applied to validate the proposed system~\cite{wong2015performance}.
In each 5-fold cross-validation, the 523 speech samples were randomly divided into five equal partitions and each partition served as the testing fold used for evaluation in turns.
This approach can reduce the bias (resulting from the environment) in evaluating the system.
In Section~\ref{subsec:exp_all}, all scores reported were the averages of the 10$\times$5 testing folds.
The significance of the performance between different approaches was statistically measured on the 50 testing folds using independent t-test at 95\% ($p<0.05$).
The voice samples in the testing fold were inferred under both the source domain in the clinic and the target domain corrupted by noises of A/Cs and streets to assess the effect of our system.
That is, the target domain data are a corrupted versions of the source domain data during the testing phase.
The approving institution of this study is Far Eastern Memorial Hospital, under the IRB/ethics board protocol number: 109063-E, and the date of approval was May 10, 2020.

In the target domain, we considered the two types of noises (A/C and street) at the same time.
For the 30 voice samples used for unsupervised domain adaptation during training, half of the samples were corrupted with noises of A/Cs;
the remaining 15 samples were corrupted with noises of streets.
During the testing phase, each testing fold (including 104-105 voice samples) was corrupted using the same rule.
Half of the samples (about 50) were corrupted with noises of A/Cs;
another half (about 50) were corrupted with noises of streets.
The magnitudes of noises were totally distinct between the training and testing phases, with signal-to-noise ratios (SNRs) of 0, 5 and 10 dB for training and 3, 6 and 9 dB for testing.
Here, it should be emphasized that we computed the UAR in the target domain over all 104-105 voice samples.
In other words, the target data UAR is the average score of the two types of noises.

Prior to training, the raw waves were first down-sampled from 44.1 kHz to 16 kHz and subsequently converted into LPS features using a Hamming filter, with a 31.25 ms window size and half of the window size as the frame shift.
The LPS features were normalized by the standard score before fed to the models.
During training, random segments of 127 frames (2 seconds) from the normalized LPS features were chosen as inputs for every epoch to increase the training variety of the models.
During testing, the first 2 seconds of the normalized LPS features were fixed as the inputs.

Fig~\ref{fig:input} visualizes the speech utterances (the vowel /:a/ sound) involved in the experiments.
In the first and third rows of Fig.~\ref{fig:input}, the left, center, and right columns show the paired waveform and spectrogram plots of health, neoplasm, and benign structural diseases voice signals recorded under a clean condition, respectively. In the second and fourth rows of Fig.~\ref{fig:input}, the left, center, and right columns, demonstrate the paired waveform and spectrogram plots of health, neoplasm, and benign structural diseases voice signals under two noisy conditions (A/Cs and streets), respectively.

\begin{table*}[htb]
    \caption{Comparison of SepConv-dat with its variants for domain adaptation. * indicates a significant difference (p-value < 0.05) between SepConv-dat and other variants.}
    \label{tab:overall}
    \centering
    \resizebox{2\columnwidth}{!}{
    \begin{tabular}{ccccccccc}
        \toprule
        \multirow{2}{*}{Model} & \multicolumn{4}{c}{Source Domain} & \multicolumn{4}{c}{Target Domain} \\
        \cmidrule(r){2-5}
        \cmidrule(r){6-9}
        & Health & Neoplasm & Structural & \textbf{UAR} & Health & Neoplasm & Structural & \textbf{UAR} \\
        \cmidrule(r){2-5}
        \cmidrule(r){6-9}
        StdConv     & 0.92 & 0.85 & 0.82 & 0.87* & 0.48 & 0.79 & 0.61 & 0.63* \\
        SepConv     & 0.88 & 0.85 & 0.80 & 0.85* & 0.39 & 0.78 & 0.60 & 0.59* \\
        \cmidrule(r){2-5}
        \cmidrule(r){6-9}
        SepConv-tgt & 0.46 & 0.38 & 0.71 & 0.52* & 0.64 & 0.69 & 0.56 & 0.63* \\
        SepConv-ft  & 0.78 & 0.70 & 0.67 & 0.72* & 0.74 & 0.84 & 0.60 & 0.72  \\
        SepConv-jnt & 0.85 & 0.79 & 0.74 & 0.79  & 0.69 & 0.74 & 0.65 & 0.69* \\
        \cmidrule(r){2-5}
        \cmidrule(r){6-9}
        SepConv-mmd & 0.87 & 0.76 & 0.73 & 0.79  & 0.66 & 0.84 & 0.61 & 0.70* \\
        \cmidrule(r){2-5}
        \cmidrule(r){6-9}
        SepConv-dat & 0.88 & 0.79 & 0.72 & 0.80  & 0.70 & 0.81 & 0.64 & 0.72  \\
    \bottomrule
    \end{tabular}}
\end{table*}

\begin{table}[htb]
    \caption{Usage of resources for replacing standard convolutions with separable convolutions.}
    \label{tab:compress}
    \centering
    \begin{tabular}{ccc}
    \toprule
        Model & \# Parameters ($\times 10^{3}$) & \# MACs ($\times 10^{6}$) \\
         \cmidrule(r){1-3}
        StdConv & 10.29 & 15.82 \\
        SepConv & 2.69  & 3.64  \\
    \bottomrule
    \end{tabular}
\end{table}

First, by comparing the plots of clean utterances, we can observe that health, neoplasm, and benign structural diseases sounds exhibit very different waveform-domain and time-frequency properties. 
Accordingly, we believe that a deep learning model can effectively classify these three types of sounds.
Next, by comparing the plots of clean utterances and noisy utterances, we can clearly note that the noises of A/Cs corrupted the detailed structures of the voice signals, especially in regions below 400 Hz.
In addition to the low frequencies, the noises of streets also influenced the details of high frequencies.
From the noisy waveform and spectrogram plots in Fig.~\ref{fig:input}, we can infer that voice disorder classification is more challenging since the key structural details of voice signals have been considerably covered by the noise signals.
Moreover, the obvious differences occurred not only between the clean and noisy plots, but also between the plots of the distinct noise types.
This is the reason why the domain classifier was designed to identify the A/C and street domains separately in the proposed system.

\subsection{Results}\label{subsec:exp_all}
Table~\ref{tab:overall} lists the overall performances of the various baselines and the proposed method.
First, we verified the effectiveness of the separable convolutional layers, \textbf{SepConv}. \textbf{SepConv} is similar to the architecture mentioned in Section~\ref{subsec:architecture} with the domain classifier removed and only source domain data were used during training.
\textbf{StdConv}, on the other hand, replaces the separable convolutional layers in \textbf{SepConv} with standard layers such that the arguments of the input channels, output channels, kernel size, etc. are identical in these two baselines.
Table~\ref{tab:overall} shows that the degradation is insignificant in \textbf{SepConv} with UARs reduced by only 4\% in the source domain and 2\% in the target domain when compared to \textbf{StdConv}.
With almost no dropping performance in the UARs, \textbf{SepConv} significantly reduced the model size and computational cost by 73.9\% in the number of parameters and 77.0\% in the number of MACs, as presented in Table~\ref{tab:compress}.
Because a MAC is the basic arithmetic unit of operation a model performs, counting the number of MACs in one forward-pass prediction of one input datum, which is independent of the hardware and platforms used, is one of the most common and fair approaches for comparing the computational cost. 
Otherwise, The computation time may vary when different computing hardware is used.
In the following domain adaptation experiments, \textbf{SepConv} was the basis for comparison.
Besides, the two baseline scores also showed that the noises tend to cause the models to misjudge the noisy inputs as neoplasms without domain adaptation.

Our proposed system, based on \textbf{SepConv} with 30 target domain samples provided in the DAT, is denoted by \textbf{SepConv-dat}.
Three other "supervised" variants (\textbf{SepConv-tgt}, \textbf{SepConv-ft} and \textbf{SepConv-jnt}) and one "unsupervised" variant (\textbf{SepConv-mmd}) were constructed for systematic comparison, with the architecture fixed as \textbf{SepConv} yet the training strategies slightly altered as follows:
\begin{itemize}
    \item \textbf{SepConv-tgt}: The model was trained \textit{only on the 30 target domain samples with labels.}
    In turn, the target domain was an exposed domain to \textbf{SepConv-tgt}, yet the source domain became unseen.
    \item \textbf{SepConv-ft}: This variation used a pretrained \textbf{SepConv} as an initial state and was then fine-tuned using the 30 target samples with labels.
    \item \textbf{SepConv-jnt}: \textbf{SepConv} was trained from scratch with labeled data from both domains jointly.
    \item \textbf{SepConv-mmd}: MMD served as an unsupervised variant using statistical-based domain adaptation.
\end{itemize}

It is observed that \textbf{SepConv-dat} outperforms all baselines in the target data UAR, particularly the other systems \textbf{SepConv-tgt}, \textbf{SepConv-ft}, and \textbf{SepConv-jnt} with extra labels of the target data provided.
Compared to \textbf{SepConv}, the UAR increased by 13\% in the target data, with only slight degradation in the source domain, maintaining 80\%. 

In \textbf{SepConv-tgt}, there is no doubt that the target data UAR is better than that of the source data, due to the exposed target domain.
However, the UAR of \textbf{SepConv-tgt} in the exposed (target) domain was not comparable with that of \textbf{SepConv} in the exposed (source) domain, with a reduction of up to 22\%.
Moreover, the UAR in the exposed (target) domain was even worse than that of the other compared systems in the unseen (target) domain.
The poor performance reflects the impact of the small dataset.

The fine-tuning of \textbf{SepConv-ft} successfully improved the UAR in the target domain by 13\%, but the degradation in the source domain was also obvious.
The average UAR of \textbf{SepConv-ft} is almost equal to that of \textbf{SepConv}, which means that we simply obtained a trade-off between the two domains.
The generalizability of model was not fully achieved by fine-tuning the target domain.

Intuitively, \textbf{SepConv-jnt} should achieve the best performance with sufficient data in both domains.
However, owing to the extremely small amount of target data under the proposed scenario, a severe imbalance of the two domains confines the improvement of generalizability.
Even the target data become distractions that degrade the source domain score.
Therefore, for the scenario with severe data imbalance that we intend to overcome, unsupervised domain adaptation algorithms are more suitable than supervised ones.

Compared to \textbf{SepConv-jnt}, \textbf{SepConv-mmd} reduces the effect of data imbalance by computing the distance between the means of the extracted features across domains.
However, obtaining statistical values that can reflect the entire target domain through only 30 target data points is almost impossible.
Finally, the total scores of \textbf{SepConv-mmd} were still worse than our proposed \textbf{SepConv-dat}.
Consequently, \textbf{SepConv-dat} is the best method among these variants, which yields the largest improvement in the target domain by overcoming data imbalance due to the extremely deficient target data.
Meanwhile, the high performance maintained in the source domain validates the generalizability.

Furthermore, we can observe that \textbf{SepConv-dat} is significantly different between the baselines and variants in most UARs, except for the target data UAR of \textbf{SepConv-ft};
the source data UAR of \textbf{SepConv-jnt} and \textbf{SepConv-mmd}.
From the results, we first note that \textbf{SepConv-ft} specializes in the performance of the target domain after fine-tuning, and our system still achieves a comparable and similar target data UAR.
Second, the labels of the source domains are exposed to \textbf{SepConv-dat}, \textbf{SepConv-mmd}, and \textbf{SepConv-jnt}, thus, they achieve a rather stable source data UAR.
In conclusion, these three scores without significant differences happen to be the strengths of the corresponding variants.
This confirms the power of \textbf{SepConv-dat}.

\subsection{Ablation: impact of domain classifier via $\lambda$}
\begin{figure}[htb]
    \centering
    \includegraphics[width=\columnwidth]{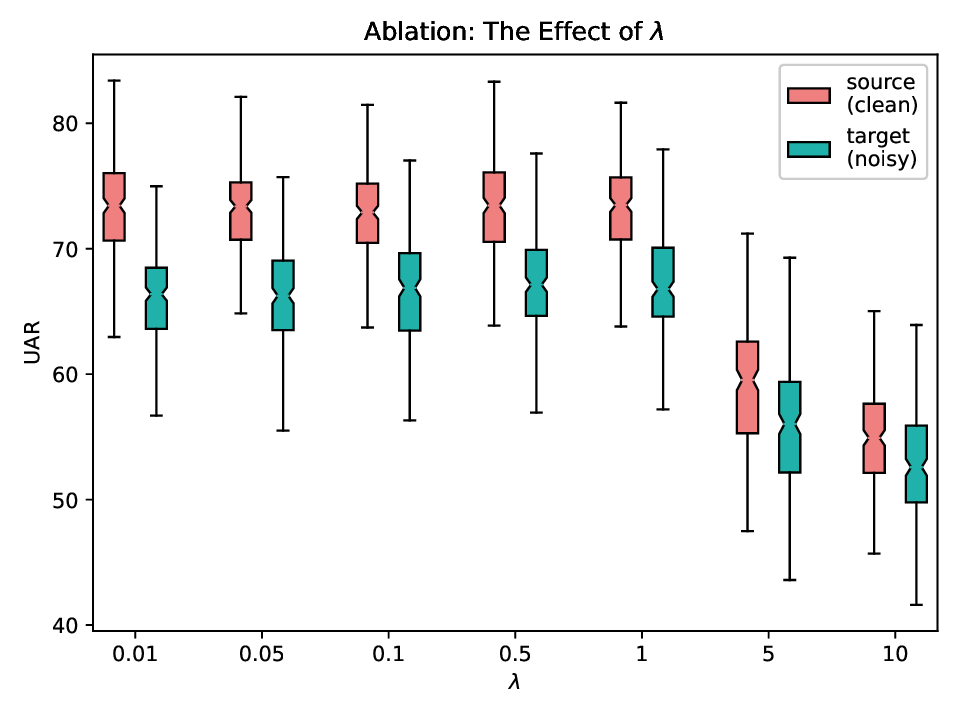}
    \caption{Box plots of different $\lambda$. The source domain is the clean data collected in the clinics, and the target domain is the data corrupted by the noises of A/Cs and streets.}
    \label{fig:lambda}
\end{figure}

We conducted an ablation study to learn how the regularization coefficient $\lambda$ in Eq.~(\ref{eq:extractor_grad}) and (\ref{eq:classifier_grad}) affect the domain adaptation performance.
Owing to the unstable training procedure of the adversarial min-max method, the statistics of 200 models with random initialized weights were considered for each setup in the ablation experiment.
Fig.~\ref{fig:lambda} only shows the results of a specific fold, but the other folds are similar.

\begin{figure}[htb]
    \centering
    \includegraphics[width=\columnwidth]{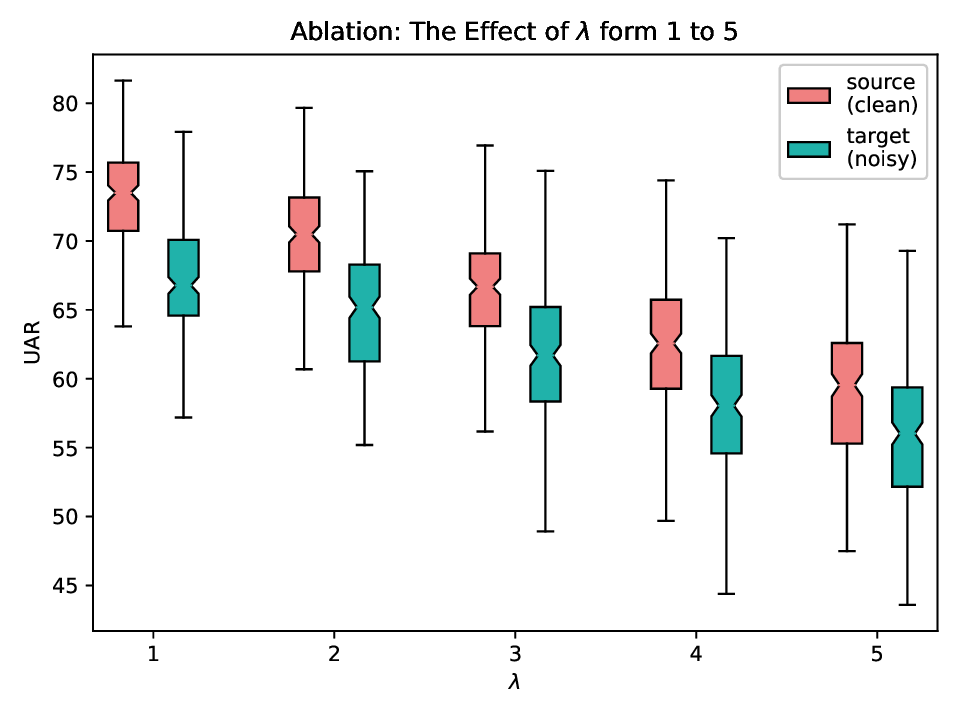}
    \caption{The detailed process of $\lambda$ increasing from 1 to 5.}
    \label{fig:lambda_1to5}
\end{figure}

The coefficient $\lambda$ was tuned from 0.01 to 10 to understand the effect.
The results in Fig.~\ref{fig:lambda} indicate that when $\lambda \geq 5$, the UAR performances in both domains broke down promptly.
It was due to the gradient of $L_d$ being over-amplified by the large $\lambda$, such that the overall update was guided away from the direction of minimizing $L_y$ to affect the diagnosis prediction.
The detailed process of increasing  $\lambda$ from 1 to 5 exhibited in Fig.~\ref{fig:lambda_1to5} corroborates our inference:
With the increment of $\lambda$, the classification gradually got worse.

On the other hand, the observed fact that when $\lambda \to 0$, the less domain-invariant the features are meets our intuition.
The coefficient is somewhere between $[0, 5)$ to best construct the domain-invariant features.
In this study, $\lambda \leq 0.5$ was observed to yield converging UARs in the source data, so that $\lambda=0.5$ was eventually chosen to balance the accuracy and the domain invariance.

\subsection{Ablation: reduction of performance in source domain}
To create the domain-invariant features, DAT guides the feature extractor to adjust in the opposite direction of minimizing the domain classifier loss $L_d$.
Nevertheless, this approach inevitably diverges from the direction of minimizing the label predictor loss $L_y$, leading to a reduction in the UAR in the source domain (clean environment).

In this section, we propose a potential solution, a two-stage inference system, to alleviate the reduction of performance in the source domain.
We introduced a noise detector that can identify whether the input voice samples are corrupted by noise or noise-free.
In the first stage, the noise detector categorizes input voice samples as either corrupted by noise or noise-free.
In the second stage, if the input voice samples are detected as corrupted by noise, we adopt the proposed \textbf{SepCNN-dat}, whereas if they are identified as noise-free, we employ the original \textbf{SepConv}.

In our experiment, we utilized a noise detector formed by CNN with the output layer having two dimensions.
The presented results are the average of 10$\times$5 testing folds.
The noise detector showed an impressive classification accuracy of 91$\%$ in the first stage of the inference process. Overall, the two-stage inference system achieved UARs of 0.84 and 0.71 in the source and target domains, respectively, which is very close to the best result of the single model (0.85 for SepConv and 0.71 for SepConv-dat).
The results of the two-stage inference system are provided in Table~\ref{tab:two-stage}.

\begin{table}[htb]
    \caption{The performance of the proposed two-stage inference.}
    \label{tab:two-stage}
    \centering
    \begin{tabular}{ccc}
    \toprule
        Model & Source UAR & Target UAR \\
        \cmidrule(r){1-3}
        SepConv     & 0.85 & 0.59 \\
        SepConv-dat & 0.80 & 0.72 \\
        \cmidrule(r){1-3}
        Two-Stage   & 0.84 & 0.71 \\
    \bottomrule
    \end{tabular}
\end{table}

\subsection{Visualization: domain invariance of extracted features}
\begin{figure}[h!]
    \centering
    \includegraphics[width=\columnwidth]{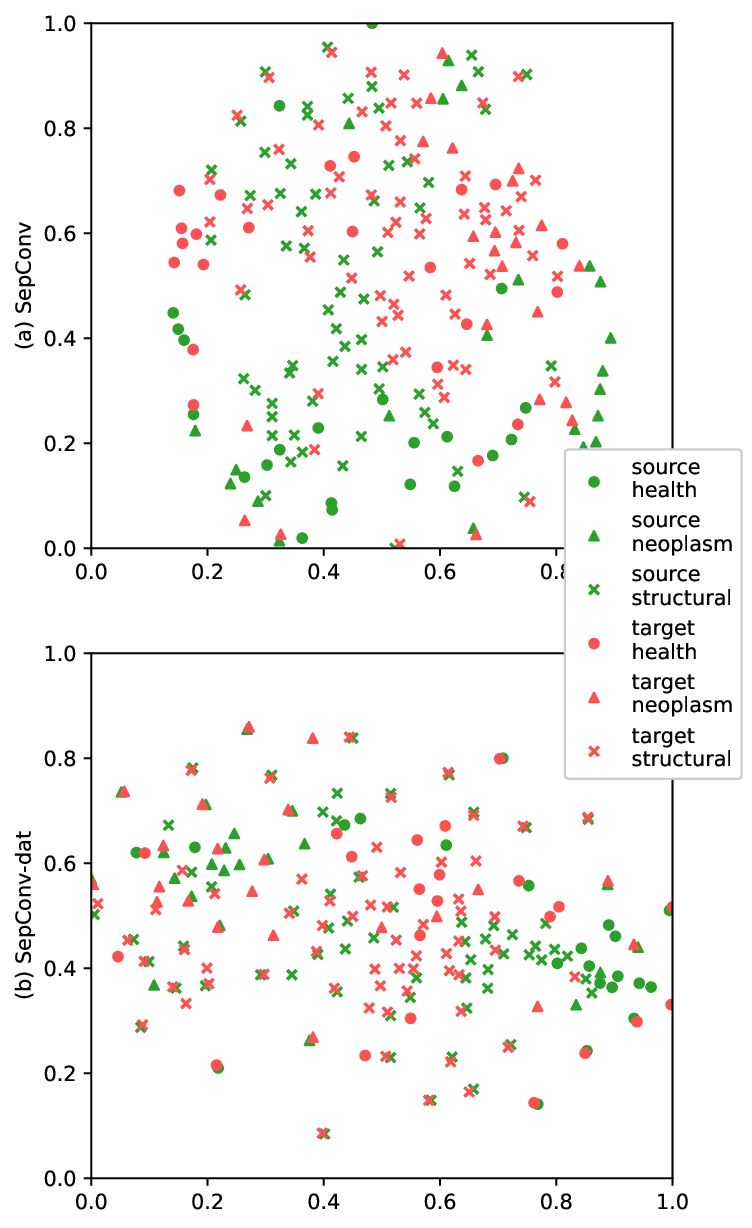}
    \caption{The t-SNE plots of the latent features extracted by the feature extractor in (a) the \textbf{SepConv} and (b) the \textbf{SepConv-dat}. The data in the source domain are marked in green, whereas the data in the target domain are marked in red.}
    \label{fig:tsne}
\end{figure}

In addition to the significant progress of the UAR in the target domain shown in Table~\ref{tab:overall}, the visualization of the distributions of features extracted from the feature extractor further proves the effectiveness of the proposed system.
Because the testing samples in the target domain are the same as those in the source domain except for the corruption by the noises, each source-target pair of extracted features should be close if high domain-invariant features are extracted.
In Fig.~\ref{fig:tsne}, the t-SNE is used to visualize the distributions of the extracted features.
Fig.~\ref{fig:tsne}(a) is the t-SNE of \textbf{SepConv} for a specified fold, but the other folds are similar.
When investigating each category in Fig.~\ref{fig:tsne}(a), these two distributions are different and have no correlation.
This explains the low accuracy of the target domain data for \textbf{SepConv}.
However, in Fig.~\ref{fig:tsne}(b), the t-SNE of \textbf{SepConv-dat} exhibits that most samples in the source domain and the target domain are in pairs. 
Whether the samples are corrupted with the A/C noises or the street noises, the proposed system could map them to corresponding clean features successfully.

%% file: 5-future.tex
\section{Future works}\label{sec:future}

We consider our future work from two perspectives.
First, in the aspect of the clinic, we are planning to perform both internal and external validations of our proposed system.
The internal validation will verify the proposed system on pathological data collected within Far Eastern Memorial Hospital (FEMH), the approval institution of this study.
Conversely, the external validation will be conducted in partnership with other hospitals to test the robustness of our system in recording environments that are unseen in this study.
In fact, We have already begun internal validation under noisy conditions, and interim results are included in the Appendix.

Second, we are also devoted to technological improvement.
Although this study investigated a more practical real-world application and achieved significant progress, the functionality of adapting to hardware mismatch should be incorporated into the proposed system, especially if implementing our approach with IoT technology or evolving it to personalized healthcare.
Therefore, our next step in technological improvement will introduce two domain classifiers, one for background noises and another for recording devices.
The interaction effects of the two min-max objective functions make the training procedure more challenging.
However, this integration allows our system to be more applicable in practice.

%% file: 6-conclusion.tex
\section{Conclusions}\label{sec:conclusion}

In the past decade, the automatic detection and classification of pathological voices have achieved outstanding performance with the advancement of machine learning methods.
Nevertheless, two main challenges arise in practical applications:
(1) state-of-the-art models often require increasing memory load and computational cost, whereas resources are rather limited for embedded systems; and (2) the domain mismatch between the training and real-world data significantly degrades the classification performance.
To overcome these difficulties, we utilized separable convolutional layers and a DAT module to build a compressed and domain-robust system.
Seven experiments were conducted and their results were compared.
The effect of $\lambda$ was also discussed.
Therefore, We proposed an unsupervised domain adaptation system that is jointly trained by using sufficient labeled data in the source domain and a small amount of unlabeled data in the target domain.
The results showed that the UAR in the noisy real-world domain improved by 13\%, and that in the clinic domain remained at 80\% with only slight degradation.
Moreover, the numbers of parameters and MACs were significantly reduced by 73.9\% and 77.0\%, respectively.

It is concluded that our proposed system efficiently reduces computational and memory usage, and effectively eliminates the domain mismatch.

%% file: appendix.tex
\appendix
We gathered voice data from 21 participants, with 10 having healthy voices and 11 having benign structural diseases (unfortunately, we were unable to obtain voice samples from patients with neoplasms due to limited revision time).
All of the voice samples we collected were corrupted by air conditioning noise in the real world.

Table~\ref{tab:real} shows the experimental results of \textbf{SepConv} and \textbf{SepConv-dat} tested with real-world and synthetic data.
As neoplasm data was not available, the UARs are the average over the recall score of healthy and benign structural diseases.
From the table, it can be observed that the prediction results of both \textbf{SepConv} and \textbf{SepConv-dat} are affected when the input is changed from synthetic data to real-world data due to acoustic mismatches.
Notably, SepConv-dat, which employs DAT techniques, still exhibits better performance than \textbf{SepConv}.
This validates that despite being trained with synthetic noisy data, \textbf{SepConv-dat} still provides notable performance improvements against noise in a real-world scenario.

It is also noteworthy that \textbf{SepConv-dat} exhibits only a 5\% reduction in UAR, while the original \textbf{SepConv} displays a decrease of 12\% in UAR.
If we regard the real-world data as an unseen domain during the training phase, \textbf{SepConv-dat} again provides better generalization ability when encountering unknown domains.

\begin{table}[htb]
    \caption{Experimental results of \textbf{SepConv} and \textbf{SepConv-dat} tested with real-world and synthetic data}
    \label{tab:real}
    \centering
    \resizebox{\columnwidth}{!}{
    \begin{tabular}{ccccccc}
        \toprule
        \multirow{2}{*}{Model} & \multicolumn{3}{c}{Real-world Data} & \multicolumn{3}{c}{Synthetic Data} \\
        \cmidrule(r){2-4}
        \cmidrule(r){5-7}
        & Health & Structural & \textbf{UAR} & Health & Structural & \textbf{UAR} \\
        \cmidrule(r){2-4}
        \cmidrule(r){5-7}
        SepConv     & 0.30 & 0.45 & 0.38 & 0.39 & 0.60 & 0.50 \\
        \cmidrule(r){2-4}
        \cmidrule(r){5-7}
        SepConv-dat & 0.60 & 0.63 & 0.62 & 0.70  & 0.64 & 0.67 \\
        \bottomrule
    \end{tabular}}
\end{table}